\documentclass[reprint,prl,twocolumn,superscriptaddress,showpacs,nofootinbib,floatfix]{revtex4-1}

\usepackage[dvips]{graphicx}
\usepackage{epsfig,amsmath,amssymb,verbatim,mathrsfs,array,layout,textcomp,latexsym,slashed,graphicx,bbm}

\usepackage[normalem]{ulem}
\usepackage{appendix}

\usepackage{rotating}
\usepackage{hyperref}
\usepackage[usenames,dvipsnames]{color}

\usepackage{soul}

\newcommand{\en}{\epsilon}
\newcommand{\bi}{\begin{itemize}}
\newcommand{\ei}{\end{itemize}}

\newcommand{\p}{\partial}
\newcommand{\mt}{\mathtt}

\newcommand{\be}{\begin{equation}}
\newcommand{\ee}{\end{equation}}
\newcommand{\bea}{\begin{eqnarray}}
\newcommand{\eea}{\end{eqnarray}}

\definecolor{gbcolor}{rgb}{.8,.3,.1}

\usepackage[utf8]{inputenc}
\usepackage[english]{babel}
\usepackage{color}

\def\beq{\begin{equation}}
\def\eeq{\end{equation}}

\begin{document}


\title{Non-local Non-Abelian Gauge Theory: \\ \it{ Conformal Invariance} \& $\beta$-function}

\author{Anish Ghoshal}
\affiliation{INFN - Sezione Roma “Tor Vergata”,Via della Ricerca Scientifica 1, 00133, Roma, Italy}
\author{Anupam Mazumdar}
\affiliation{Van Swinderen Institute, University of Groningen, 9747 AG Groningen, The Netherlands}
\author{Nobuchika Okada}
\affiliation{Department of Physics and Astronomy, University of Alabama, Tuscaloosa, Alabama 35487, USA}
\author{Desmond Villalba}
\affiliation{Department of Chemistry and Physics,
Drury University, Springfield , Mo 65802, USA}

\begin{abstract}
This paper focuses on extending our previous discussion of an Abelian U(1) gauge theory involving infinite derivatives to a non-Abelian SU(N) case. The renormalization group equation (RGEs) of the SU(N) gauge coupling is calculated and shown to reproduce the local theory 
$\beta$-function in the limit of the non-local scale M $\rightarrow \infty$. 
Interestingly, the gauge coupling stops its running beyond the scale $M$, approaching an asymptotically conformal theory. 

\end{abstract}

\maketitle

\section{Introduction}
\label{Intro}

With no clinching evidence of new particles in physics Beyond the Standard Model (BSM) by any of the current searches at experiments world-wide, such as the Large Hadron Collider (LHC), an alternative philosophy for BSM could be the modification of the standard canonical kinetic terms through the introduction of infinite derivatives, instead of introducing new particles (new states). Motivated by string field theory \cite{sft1,sft2,sft3,padic1,padic2,padic3,Frampton-padic,Tseytlin:1995uq,marc,Siegel:2003vt}, infinite derivative formulation is expressed in the form of an entire function \cite{Biswas:2014yia} and the higher order derivatives are  accompanied by a suppression by a scale $M$, 
which we call the ``non-local scale.''  The choice of this derivative function does not appear to be unique, as long as it acts to suppress terms
in the high energy regime.

We have previously considered
a non-local Abelian U(1) gauge theory within this framework and have shown that the evolution of the gauge coupling becomes fixed or ``UV-insensitive'' 
in the energy regime well above the non-local scale $M$ \cite{Ghoshal:2017egr}.
It was also shown that Higgs vacuum instability problem \cite{Olive:2016xmw} is cured, leading to a stable Abelian Higgs theory.  The theory is ghost-free 
\cite{Buoninfante:2018mre} and predicts a unique scattering phenomenology leading to transmutation of energy scales which has its own cosmological and astrophysical implications \cite{Buoninfante:2018gce}. The phenomenology of dark matter in this theory is also investigated in Ref. \cite{Ghoshal:2018gpq} and shown DM experiments can be a novel probe for the scale on non-locality. Strongly coupled regime of the theory was considered in Refs. \cite{Frasca:2020jbe,Frasca:2021guf} in Higgs and Yang-Mills versions and it was found that the mass gap generated gets diluted due to non-local effects.
On aspects of gravity, 
Ref.~\cite{Biswas:2011ar} showed that the most general quadratic curvature gravitational action (parity-invariant and torsion-free) with infinite covariant derivatives makes the gravitational sector free from the Weyl {\it ghost} and is devoid of any classical singularities, such as black hole \cite{Biswas:2011ar,Biswas:2013cha,Frolov:2015bia,Frolov:2015usa,Koshelev:2018hpt,Koshelev:2017bxd,Buoninfante:2018xiw,Cornell:2017irh,Buoninfante:2018rlq,
Buoninfante:2018stt}~\footnote{For previous arguments related to non-singular solutions,see Refs.~\cite{Tseytlin:1995uq,Siegel:2003vt}.} and cosmological singularities~\cite{Biswas:2005qr,Biswas:2006bs,Biswas:2010zk,Biswas:2012bp,Koshelev:2012qn,Koshelev:2018rau}~\footnote{For supersymmetric versions on non-locality in the matter section, see for instance, Refs. \cite{Gama:2017ets,Gama:2020pte}}.

In this paper, we extend the same idea involving infinite series of higher-order derivatives to a non-Abelian SU(N) gauge theory, 
investigate the gauge invariance, and compute the running of the SU(N) gauge coupling. 
Here too we expect that within this framework, the standard renormalization group equations (RGEs) should be reproduced
in the local limit ($M \to \infty$).

\section{Non-Abelian Extension}
For local non-Abelian SU(N) gauge theory, the Lagrangian includes the gauge boson kinetic term, 
\be
\mathcal{L}_{g}=-\frac{1}{2} tr [F^{a\mu\nu} F_{a\mu\nu}]=
-\frac{1}{4}  F^{a\mu\nu} F_{a\mu\nu}.
\ee
The trace is over the SU(N) group indices and the field-strength tensor is given by
\be
F_{\mu\nu}^a=\p_{[\mu}A_{\nu]}^a -gf^{abc}A^b_{\mu}A^c_{\nu}     \ ,
\ee
where the $f^{abc}$ represents the group structure constant. 
For implementation of the non-local modification, we follow our approach in Ref.~\cite{Ghoshal:2017egr}. 
The gauge boson kinetic term is then described as
\be 
\mathcal{L}_{g}=-\frac{1}{2} tr [F^{a\mu\nu} e^{-{D^2 \over M^2}} F_{a\mu\nu}] + h.c. ,
\ee
where the covariant derivative is given by $D_\mu=\partial_\mu-igT^{a}A^a_{\mu}$. The fermionic part of the Lagrangian is given by the standard form as in Refs. \cite{Biswas:2014yia,Ghoshal:2017egr}:
\begin{eqnarray}
 \mathcal{L} &=& \bar{\psi} e^{{D^2 \over M^2}}   i \gamma ^\mu D_{\mu}   \psi  
\end{eqnarray}
where $D^2= \eta_{\mu\nu} D^{\mu} D^{\nu}$ $(\mu, \nu=0,1,2,3$), assuming all gauge and fermionic particles being massless.
  
The exponential term is introduced by the non-local modification and the Lagrangian includes
an infinite series of higher dimensional operators that are all suppressed by the non-local scale $M$. 
As a result, their contribution can be largely ignored at energies lower than $M$. 
In other words, the conventional Lagrangian is reproduced in the limit of $M \to \infty$. 
We take the metric with $\eta={\rm diag}(+1,-1,-1,-1)$ to implement our procedure for UV completion upon the Wick rotation.
%

\subsection{Non-Abelian Gauge Field Propagator}
To find the massless non-Abelian gauge boson propagator, we follow the same gauge-fixing prescription
as in Ref.~\cite{Biswas:2014yia}, and the non-Abelian gauge and ghost Lagrangians are given by

\be
\mathcal{L}_{ghost}=-\bar{c}^af(\Box)(\partial^\mu D_{\mu}^{ab})c^b,
\label{NLghost}
\ee
and
\begin{widetext}
\begin{equation}
\mathcal{L}_g = \frac{1}{2} A_{\mu}^{a}e^{-{\Box\over M^2}} (\Box\, \eta^{\mu\nu}-\p^{\mu}\p^{\nu})A_{\nu}^{a} +{1\over {2 \xi}}A_{\mu}^a(f(\Box))^2\p^{\mu}\p^{\nu}A_{\nu}^{a}, 
\label{NLgauge}
\end{equation}
\end{widetext}
where $\xi$ is the gauge fixing parameter and $D_\mu^{ab}=\partial_\mu \delta^{ab}-igA^c_{\mu}(T^{c})^{ab}$
is the covariant derivative in the adjoint representation. 
In order to have consistency with the standard gauge fixing procedure, 
we choose the entire function $f(\Box)=e^{-{\Box \over 2M^2}}$. 
The non-local Faddeev-Popov procedure is given in Appendix. 
In the Euclidean space, the gauge boson and ghost propagators have the following forms: 
\bea
&&\Pi_g(p^2)={i\eta_{\mu\nu}\delta^{ab}e^{-{p^2 \over M^2}}\over p^2 +i\en} , \nonumber \\
&&\Pi_{ghost}(p^2)={i\delta^{ab}e^{-{p^2 \over {2M^2}}}\over p^2 +i\en},
\label{NL-propagator}
\eea
in the Feynman-'t Hooft gauge $\xi=1$. And the massless fermion propagator is given by~\cite{Biswas:2014yia}:
\begin{equation}
\Pi_\psi(p_E) = -{i \slashed{p}_E e^{-{p_E^2 \over M^2}}\over p_E^2 + i \epsilon }\,.
\end{equation}

\section{Gauge coupling running}
In Ref.~\cite{Ghoshal:2017egr}, we have obtained the RGE for the gauge coupling in the Abelian U(1) gauge theory 
with one Dirac fermion having a unit U(1) charge: 
\bea \label{f1}
\mu\frac{dg}{d\mu}=\frac{1}{16\pi^{2}} \left(\frac{4}{3}\right)  g^3 e^{-2 \frac{\mu^2}{M^2}}. 
\eea
The standard result for the beta function is obtained in the local limit of $M \to \infty$. 
Interestingly, for the non-local U(1) theory is ``UV-complete,'' as the beta function is vanishing beyond the non-local scale. 
This behavior is opposed to the local U(1) theory where the running is asymptotically non-free. 
In the case of a non-Abelian theory, we expect a similar behavior, namely, 
the gauge coupling stops running beyond the non-local scale.
The beta function for the non-Abelian gauge coupling incorporates the gauge and ghost field contributions; we break up the derivation of each contribution below for clarity.

\subsection{Gauge Wave Function Renormalization}
First we consider the wave function renormalization for the gauge fields. We use the standard group theory parameterizations: 
$f^{acd}f^{bcd}=T(A) \delta^{ab}$ and ${\rm Tr}[T^aT^b ]=T(R)\delta^{ab}$ 
with the Dynkin indices for the adjoint ($T(A)=N$) and the fundanmental ($T(R)=1/2$) 
representations, respectively. 
The quadratic Casimir for the adjoint representatio is $C_2(A)=N$.
All the following calculations are performed with the Feynman gauge ($\xi=1$). 

Our methodology for calculating the wave function renormalization follows
from the same procedure in Ref.~\cite{Ghoshal:2017egr}. In our calculation, we employ the cutoff regularization scheme. Although one should use a gauge invariant regularization scheme such as the dimensional regularization, 
the cutoff regularization scheme can be practically used, since there is a one-to-one correspondence
between the two schemes (at the one-loop level), $\log \Lambda^2 = \frac{1}{\epsilon}$, 
where $\Lambda$ is the cutoff parameter and $\epsilon$ is the dimensional regularization parameter.
For the gauge-gauge-gauge contribution:
\bea 
\nonumber \Gamma_{2} &=&
\frac{-g^2T(A)\delta^{ab}}{2}\int \frac{d^4 p}{(2 \pi)^4}\frac{N^{\mu \nu}}{p^2(p+q)^2}e^{-\beta(p)^2}e^{-\beta(p+q)^2},
\eea
where $\beta \equiv 1/M^2$, and $N^{\mu \nu}$ is given by
\bea
&&N^{\mu \nu}=\left[g^{\mu \rho}(q-p)^\sigma+g^{\rho \sigma}(2p+q)+g^{\sigma \mu}(-p-2q)^{\rho} \right] \nonumber \\
&&\times \left[\delta^{\nu}_\rho(p-q)_{\sigma}+g_{\rho \sigma}(-2p-q)^\nu +\delta^{\nu}_{\sigma}(p+2q)_{\rho} \right]. 
\eea
Using the Schwinger parameters, $\alpha_1$ and $\alpha_2$, the integral is recast as
 \bea 
\nonumber \Gamma_{2} &=&
\frac{-g^2T(A)\delta^{ab}}{32\pi^2}\int p^2d p^2 \int d\alpha_1 d\alpha_2 \, N^{\mu \nu}\nonumber \\
&&\times e^{-(\beta(p^2+(p+q)^2)+\alpha_1(p+q)^2+\alpha_2 p^2)}.
\eea
It is convenient to introduce new parameters, $s:\{0,\infty\}$ and $\alpha:\{0,1\}$, 
which are defined as  $\alpha_1+\alpha_2=s$ and $\alpha=\alpha_1s$ 
(which gives $\alpha_2=s(1-\alpha)$). 
In addition, we take the momentum shift $ p \to p- \frac{\alpha s + \beta}{s+2\beta}q$ and then 
\bea
\nonumber \Gamma_{2} &=&
\frac{-g^2T(A)\delta^{ab}}{32\pi^2}\int p^2d p^2 \int_{0}^{1} d\alpha \int_{0}^{\infty} s ds \, \bar{N}^{\mu \nu} \nonumber \\
&&\times e^{-((s+2\beta)p^2+\frac{\alpha(1-\alpha)s^2+\beta s+\beta^2}{s+2\beta}q^2)}, 
\eea
where $\bar{N}^{\mu \nu}$ is given by
\bea
&&\bar{N}^{\mu \nu}=\left[g^{\mu \rho}(Aq-p)^\sigma+g^{\rho \sigma}(2p+Bq)+g^{\sigma \mu}(-p-Cq)^{\rho} \right] \nonumber \\
&&\times \left[\delta^{\nu}_\rho(p-Aq)_{\sigma}+g_{\rho \sigma}(-2p-Bq)^\nu +\delta^{\nu}_{\sigma}(p+Cq)_{\rho} \right], 
\eea
and 
$A=1+\frac{\alpha s+\beta}{s+2\beta}$, $B=1-\frac{2\alpha s+2\beta}{s+2\beta}$, and $C=2 -\frac{\alpha s+\beta}{s+2\beta}$. 
To find the corrections $\Delta Z_{Gauge}$, we focus on the coefficient of terms only proportional to $q^{\mu}q^{\nu}$, 
which gives 
\bea
\Gamma_{2} &=&
\frac{-g^2T(A)\delta^{ab}q^{\mu}q^{\nu}}{16\pi^2}\int p^2d p^2 \int_{0}^{1} d\alpha \int_{0}^{\infty} s ds \, (A(C-B) \nonumber \\&&+BC-2B^2)\times e^{-(s+2\beta)p^2}\left[1+O(q^2)\right]. 
\eea
Here, $O(q^2)$ contains higher order momentum terms which are sub-leading and ignored:
\bea\label{A1}
\Gamma ^{gauge} _{2} &\approx & \frac{-g^2T(A)\delta^{ab}q^{\mu}q^{\nu}}{16\pi^2}\int p^2dp^2 \int_{0}^{1} d\alpha \int_{0}^{\infty} s ds \,  \nonumber \\
&\times &(1+5\alpha-5\alpha^2)e^{-(s+2\beta)p^2} \nonumber \\
&\approx & \frac{-g^2T(A)\delta^{ab}q^{\mu}q^{\nu}}{16\pi^2} \left(\frac{11}{6}\right)\int_{0}^{\Lambda^2} d^2 p \, \frac{e^{-2\beta p^2}}{p^2}
\eea

Employing the same procedure for the ghost contribution, we find 
\bea
\Gamma ^{ghost}_{2} &=&
\frac{-g^2T(A)\delta^{ab}}{16\pi^2}\int p^2d p^2 \int_{0}^{1} d\alpha \int_{0}^{\infty} s ds \nonumber \\
&&\times \left(p+\left(1-\frac{\alpha s + \beta'}{s+2\beta'}\right)q\right)^{\mu}\left(p-\frac{\alpha s + \beta'}{s+2\beta'}q\right)^{\nu} \nonumber \\
&&\times e^{-(s+2\beta')p^2)}\left[1+O(q^2)\right], 
\eea
where $\beta'=\beta/2$ due to the ghost propagator's exponential factor in (\ref{F1}). 
Again, picking up the $q^{\mu}q^{\nu}$ terms and ignoring the further sub-leading terms,
\bea\label{A2}
\Gamma ^{ghost}_{2} &\approx&
\frac{g^2T(A)\delta^{ab}q^{\mu}q^{\nu}}{16\pi^2}\int p^2d p^2 \int_{0}^{1} d\alpha \int_{0}^{\infty} s ds \nonumber \\
&&\times \frac{((1-\alpha)s+\beta')(\alpha s+\beta')}{(2 \beta' + s)^2}e^{-(s+2\beta')p^2} \nonumber \\
&\approx & \frac{g^2T(A)\delta^{ab}q^{\mu}q^{\nu}}{16\pi^2} \left(\frac{1}{6}\right)\int_{0}^{\Lambda^2} d p^2 \, \frac{e^{-\beta p^2}}{p^2}.
\eea

Contributions from the fermion loops are the same as in Ref.~\cite{Ghoshal:2017egr}
with the addition of the SU(N) group factor $T(R)$, for $N_F$ fermions
\bea\label{A3}
\Gamma ^{fermion} _{2} &\approx&
\frac{g^2N_FT(R)\delta^{ab}q^{\mu}q^{\nu}}{16\pi^2} \left(\frac{4}{3}\right)\int_{0}^{\Lambda^2} dp^2 \, \frac{e^{-2\beta p^2}}{p^2}.
\eea

Extracting all the relevant terms from (\ref{A1}) and (\ref{A2}), 
\bea\label{A4}
 \Delta Z_{gauge}&&=\frac{g^2}{16 \pi ^2}\int_{0}^{\Lambda^2} \frac{d p^2}{p^2} \left\lbrace \left[-\frac{11}{6}T(A) +\frac{4}{3}N_F T(R)  
 \right]\right. \nonumber \\
&& \left. \times e^{-2\beta p^2} +  \frac{1}{6}e^{-\beta p^2}T(A) \right\rbrace.
\eea

\begin{figure}[h!] \vspace{1cm}
\centering
\includegraphics[height=6cm, width=8cm]{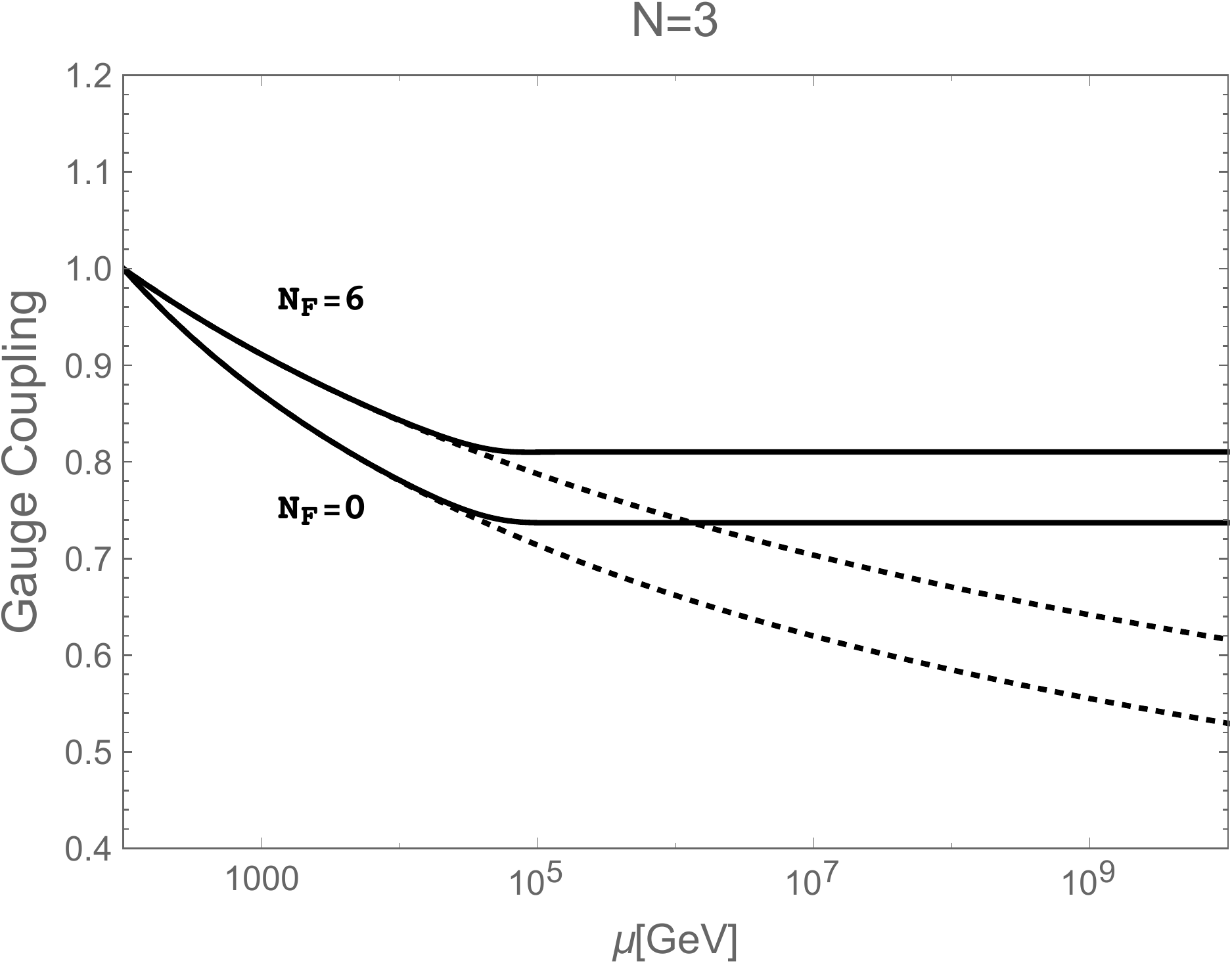} \hspace{1cm}
\caption{
The SU(3) gauge coupling running with $N_{F} =6$ \& $N_{F} =0$, shown in solid (dashed) black lines for the non-local (local) theories. 
Here, we have set $M=10^5$ GeV. }
\label{f1}
\end{figure}

\medskip

\subsection{Fermion Wavefunction Renormalization}

Proceeding in a similar manner, lastly, the non-Abelian gauge contribution to the fermion wave function renormalization is
\bea\label{A6}
\Delta Z_{fermion} =\frac{g^2C_2(R)}{16 \pi ^2}\int_{0}^{\Lambda^2} \frac{d p^2}{p^2}e^{-2\beta p^2}.
\eea


\subsection{Non-Abelian Vertex Correction }

Next we consider the corrections to the non-Abelian fermion-fermion-gauge vertex. They are evaluated in a similar fashion, with the main difference being external momentum is set to zero. We work in the Feynman-'t Hooft gauge, and find
\bea\label{A5}
\Delta g &&=\frac{g^3}{16 \pi ^2}\int_{0}^{\Lambda^2} \frac{d p^2}{p^2} \left\{ \left(C_2(R)-\frac{1}{2}T(A) \right) \right. \nonumber \\
&& \left. +  \frac{3}{2}T(A) \right\} e^{-3\beta p^2}
\eea

Having derived wave-function renormalization and vertex correction we proceed to investigate the $\beta-$function studies \footnote{In deriving the RGE, one may consider 1-loop corrections to the gauge boson self-interactions 
or 1-loop corrections to the gauge coupling of the ghost. 
Thanks to the gauge invariance, the RGE devised in this way is coincide with the one presented in this paper.}.




\subsection{Non-Local Non-Abelian Beta Function}

Bringing these contributions together in the standard way we culminate with the beta function \footnote{The RGE is described as $\delta g = \Delta g - g (\Delta Z_{fermion} + \frac{1}{2} \Delta Z_{gauge}) $} for the gauge couplings in the infinite derivative framework.

 According to the standard QFT procedure, we extract the $\beta$-function from $\Gamma_2$ by taking $\partial / \partial \rm log \Lambda$
and replacing the cutoff $\Lambda$ with the re-normalization scale $\mu$\footnote{Also see Ref.~\cite{Ghoshal:2017egr} for details of the procedure.}
.

 \begin{widetext}
 \be
\mu \frac{dg}{d\mu}=\frac{-g^3}{16 \pi ^2}\left\lbrace -\frac{1}{6}C_2(A)e^{-\beta \mu^2}
- \left( -\frac{11}{6}C_2(A) +\frac{4}{3}N_F T(R)+2C_2(R)  \right)e^{-2\beta \mu^2} + 2\left(C_2(R)+T(A)\right)e^{-3\beta \mu^2} \right\} .
\ee
\end{widetext}

In the limit $M \rightarrow \infty$ ($\beta \rightarrow 0$) we recover the standard SU(N) RGE, 
which is what we expect in the infrared (IR) limit of the non-local theory: 
\bea
\mu \frac{dg}{d\mu}&&=\frac{-g^3}{16 \pi ^2}\left\{\frac{11}{3}N - \frac{2}{3}N_{F}\right\}.
\eea
In Fig.~\ref{f1} we show the gauge coupling running of the non-local (local) non-Abelian SU(N) gauge theory 
represented by the solid (dashed) line, where we have set $N=3$ and $N_{F} =6$ \& $N_{F} =0$. 
Beyond the non-local scale that we set $M=10^5$ GeV, the running becomes ``conformally complete'', 
in the sense that the running becomes frozen. 
On the other hand, the standard gauge coupling running exhibits the usual asymptotic free behavior.


\section{Conclusions and Discussions}
\label{SI}

The central attractive feature of our non-local extension of QFT is that 
the theory becomes scale free (scale invariant) at energies beyond the non-local scale $M$.
In other words, the theory becomes conformal, and $M$ signifies the UV-fixed point \footnote{See Refs. \cite{Martin:2000cr,Bond:2017suy} for UV-fixed points in supersymmetric context.}. 
The UV behaviour of the non-Abelian Higgs model is expected to be very similar to
what we have discussed in the Abelian case \cite{Ghoshal:2017egr}.
The RG evolution of the Higgs self-coupling freezes beyond $M$ 
and the Higgs potential never develops any instability.

Classically scale-invariant models are of immense interest in QFT. Usually the conformal symmetries are anomalous in QFT in four-dimensions except for a specific system such as N=4 supersymmetric Yang-Mills theory. We may classify scale-invariant theories as: \\
(i) Exact scale invariance is always maintained; \\
(ii) Scale Invariance breaks only at the quantum level.

In infinite derivative non-Abelian gauge theory, we find a unique scenario where the theory does not possess the scale invariance in the IR, but theory becomes scale invarinat in the UV whose scale is set
by the non-local scale $M$. 
Thus, the symmetry breaking (to generate a scale) maybe considered as an artefact of only the low energy behaviour of the theory leading to the concept of scale-dependence, or in other words, 
the running of the coupling constants. 
This is very similar to the classical scale invariant theory with “soft” symmetry-breaking, 
which would not suffer from the naturalness issue arising from the UV sensitivity
of scalar mass squared corrections. 
However, in non-local theory, the beta function is exponentially suppressed for $\mu > M$, 
so that energy $M$ practically works as the conformal fixed point. 
This leads to scale invariance in the UV \textit{in spite of} “quantum" interactions and the presence of “scale" in the IR, thereby denoting a “scale-insensitivity" of the \textit{tree-level} action
\footnote{   
The idea of scale invariance in context to Higgs naturalness issue was proposed long time ago \cite{Adler:1982ri,Coleman:1973jx}, it has recently received great attention with respect to UV-complete framework to address the hierarchy problem \cite{Salvio:2014soa,Einhorn:2014gfa,Einhorn:2016mws,Einhorn:2015lzy,Bardeen,Foot:2007iy,AlexanderNunneley:2010nw,Englert:2013gz,Hambye:2013sna,Farzinnia:2013pga,Altmannshofer:2014vra,Holthausen:2013ota,Salvio:2014soa,Einhorn:2014gfa,Kannike:2015apa,Farzinnia:2015fka,Kannike:2016bny,Hiller:2019mou,Hiller:2020fbu}}. 



In the infinite derivative theory, the Higgs mass squared corrections are exponentially suppressed at the scale beyond $M$, and the non-local scale works as an effective cutoff for the corrections, $\Delta m_{H}^2 \sim M^2$ \cite{Biswas:2014yia}. This means that the Higgs mass fine-tuning is reduced to $\frac{M_{EW}^2}{M ^2}$ from the usual $\frac{M_{EW}^2}{M_{Planck}^2}$, with $M_{Planck}$ being the Planck mass.
Beyond the scale of non-locality to infinite energy, all the beta-functions are vanishing and all the couplings are approaching a fixed point, determined by the non-local scale (see Fig.~\ref{f1}). This means that \textit{no Landau-pole} exists in the theory. 

We end our discussion with the speculation that the non-local extension of gauge field theories may allow us to provide a unified framework of 
\textit{Conformal Invariance} without having to encounter
with Landau poles, and thus paves a theoretical pathway
for theories being perturbatively stable and valid to infinite energy. However, the details are beyond the scope of the current investigation, and we will take this issue up explicitly in a future publication.

%

%
%


\subsection{Appendix A: NL Non-Abelian Faddeev-Popov Procedure}
Focusing solely on the kinetic term of the non-abelian gauge field, 
the partition function in the Euclidean space is given by
\bea
\mathcal{Z}(J)&&=\int DA^a \int D\alpha^a \delta (G^a(x))det\left(\frac{\delta G^a(x)}{\delta \alpha^a}\right) \nonumber \\
\times && e^{-\int d^4x(-\frac{1}{4} F^{a\mu\nu}e^{-{\Box \over M^2}} F_{a\mu\nu})},
\eea
where $G^a(x)=f(\Box)\partial^\mu A^a_\mu(x)-w^a(x)$ and $G^a(x)$ transforms as $G^a(x) \to f(\Box)\partial^\mu (A^a_\mu(x)+D^{ab}_\mu \alpha^b(x))-w^a(x)$ under a SU(N) gauge transformation. Because the function $\mathcal{Z}(J)$ is independent of $w^a(x)$ we introduce the arbitrary functional dependent on $w^a(x)$ and employ the usual gauge fixing procedure to arrive at 
\bea
\mathcal{Z}(J)&& \propto \int Dw \int DA \int D\alpha\, \delta (G(x))det\left(f(\Box)\partial^\mu D^{ab}_\mu \delta^4(x-y)\right) \nonumber \\
\times && e^{-\int d^4x(-\frac{1}{4} F^{a\mu\nu}e^{-{\Box \over M^2}} F_{a\mu\nu}-\frac{1}{2 \xi}w^2)}.
\eea
Using the familiar relation between a functional determinant and a path integral over complex Grassmann variables, we obtain the lagrangians given in Eqs.~(\ref{NLghost}) and (\ref{NLgauge}). 
In order to provide consistency with the gauge fixing procedure in the local limit, we have chosen $f(\Box)=e^{-\Box/2M^2}$.


\subsection{Appendix B: Renormalizability and BRST Invariance}
%

\subsubsection{Power Counting}

In conventional quantum field theories, the kinetic terms of gauge fields contain up to two derivatives. In momentum space this means that the propagators behave as $k^{-2}$. In four dimensions each momentum loop provides a $k^4$ factor in any quantum loop integral. The superficial degree of divergence of a Feynman diagram in the local theory is therefore given by 
\bea
D &=& \text{no. of factors of internal momentum in the numerator} \nonumber \\
& & - \text{no. of factors in the denominator} = 4 L - 2 I + 2 V , \nonumber \,
\eea
where $L$ is the number of loops, $V$ is the number of vertices, and $I$ is the number of internal propagators.

When one has exponential factors in the loops, vertices and propagators, an exponential suppression will always be dominant over a polynomial growth. Therefore, we see that as long as these exponential factors come with a negative power, the integrals should be manifestly convergent. 
%
In the counting of the superficial degree of divergence in the infinite derivative theory, 
we note that each propagator comes with an exponential suppression, (see Eq.~(\ref{NL-propagator}) 
and Ref.~\cite{Ghoshal:2017egr}) and 
each vertex also comes with an exponential suppression. 
Thus, the power of exponential suppression factor should be:
\be
E = -V - I\,.
\ee
By using the topological relation:
\be
V = I + 1 - L,
\ee
we have
\be
\label{eq:E}
E = 1 - 2V - L \,.
\ee
Since V is an integer, $E < 0$, for all loops and the corresponding scattering amplitudes are superficially convergent. 

\medskip

\subsubsection{BRST Invariance}

Next we discuss the convergence in the complete BRST-invariant infinite-derivative gauge theory action. The conclusion is exactly the same. 
The quantized action action is of the form: 
\begin{widetext}
\bea \label{F1}
&&{\mt{S}_{\mt{total, \,quantized}}} = S_{\mt g} + S _ {\mt{Gauge-Fixing}} + S _ {\mt{ghost}} \nonumber \\
&=& \int d^4 x \left(-\frac{1}{4} (F_{\mu\nu}^a  e^{-D^2 \over M^2} (F^a) ^{\mu\nu}) + \frac{\xi}{2} (B^a) ^2 + B^a \partial ^{\mu} A_{\mu}^a + \bar{c} ^a (-\partial ^{\mu} e^{-\Box \over 2M^2} D_{\mu}^{ac})c^c \right),
\eea
\end{widetext}
where $\xi$ is the gauge fixing parameter, $B$ is the auxiliary field, and $c$ and $\bar{c}$ are the ghost and anti-ghost fields, respectively.
The BRST transformations for non-Ableian gauge theories express a residual symmetry
of the effective action which remains after the original gauge invariance has been
broken by the addition of the gauge-fixing and ghost action terms. 
In our theory, we introduce the following BRST transformations: 
\bea \label{BRST1}
\delta _ {\mt{BRST}} (A ^ {\mu })^a & = (D_{c} ^{\mu})^a c ^ {c} \delta \lambda\,,  \\ \label{BRST2}
\delta _ {\mt{BRST}} c ^ {a} & = -\frac{1}{2} g f^{a b c} c ^ {a} c ^ {b} \delta \lambda\,, \\ \label{BRST3}
\delta _ {\mt{BRST}} \bar{c} ^a & = (\delta \lambda) e ^{\Box  \over M ^2} B^a \,, \\ \label{BRST4}
\delta _ {\mt{BRST}} B^a & = 0 .
\eea
where $\delta \lambda$ is an infinitesimal anti-commuting constant parameter. 
We show the BRST-invariance of 
$S _ {\mt{total, quantized} }$ (following Refs.~\cite{Moffat:2011an,Talaganis:2014ida}):
the BRST transformation of the gauge field is just a gauge transformation of $A_{\mu }$ generated
by $c_{a} \delta \lambda$. Therefore, any gauge-invariant functionals of $F_{\mu \nu}$, 
like the first term in Eq.~(\ref{F1}) gives
$\delta _ {\mt{BRST}} (-\frac{1}{4} (F_{\mu\nu}^a  e^{-D^2 \over M^2} (F ^{\mu\nu})^a)) = 0.$
The second term in Eq.~(\ref{F1}) gives $\delta _ {\mt{BRST}} (\frac{\xi}{2} (B^a) ^2) = 0$ from Eq.~(\ref{BRST4}).
For the third term in Eq.~(\ref{F1}), the transformation of A$_{\mu} ^a$ cancels the transformation of $\bar{c}$ in the last term, due to Eqs.~(\ref{BRST1}, \ref{BRST2} \& \ref{BRST3})  leaving us with
\be \label{F2}
\delta_{\mt{BRST}} (D^{ac} _{\mu} c^c) = D^{ac} _{\mu} \delta c^c + g f^{a b c} \delta A ^b_{\mu} c^c,
\ee
 which is is equal to 0, using the \textit{Jacobi identity} (see Ref.~\cite{Peskin}).
The transformation of $c^{\sigma}$ is nilpotent,
\be
\delta_{\mt{BRST}} (\partial_{\mu} c^{a}c^{b})=0\,,
\ee
while the transformation of $A^{\mu}$ is also nilpotent,
\be
\delta_{\mt{BRST}}((D_{b}^{\mu})^a c^{b}) = 0\,.
\ee

Hence, the action in Eq.~(\ref{F1}) is BRST-invariant.
Noting the fact that the only part of the ghost action which varies under the BRST transformations is that of  the anti-ghost ($\bar{c} _{a}$), the central idea behind our proof of BRST-invariance is that 
we have chosen the BRST variation of the anti-ghost ($\bar{c} _{a}$) (see Eqs.~(\ref{BRST3}))
to cancel the variation of the gauge-fixing term.

Next we proceed in the same manner as in Ref.~\cite{Talaganis:2014ida} and 
introduce BRST-invariant couplings of the ghosts and gauge bosons to external fields
$K_{\mu}$ (anti-commuting) and $L_{\sigma}$ (commuting), 
so that the effective action $\widetilde{S}$ is
\be
\widetilde{S} = S_{\mt{total,quantized}}+ K _ {\mu } D _{a} ^ {\mu } c ^{a} + L _{\sigma} f^{a b c} c ^ {a} c ^ {b}\,,
\ee
which is also BRST-invariant.

Let us now compute the superficial degree of divergence for the BRST-invariant action. 
To proceed, we introduce the following notations:
\bi
\item $n_{A}$ is the number of gauge boson vertices,
\item $n_{G}$ is the number of anti-ghost-gauge boson-ghost vertices,
\item $n_{K}$ is the number of $K$-gauge boson-ghost vertices,
\item $n_{L}$ is the number of $L$-ghost-ghost vertices,
\item $I_{A}$ is the number of internal gauge boson propagators,
\item $I_{G}$ is the number of internal ghost propagators,
\item $E_{c}$ is the number of external ghosts,
\item $E_{\bar{c}}$ is the number of external anti-ghosts.
\ei
By counting the exponential contributions of the propagators and the vertex factors, as discussed earlier, we can now obtain the superficial 
degree of divergence, which is given by
\be
E = - n _ {A} - I _ {A}\,. 
\ee
By using the following topological relation,
\be
L = 1 + I _ {A} + I _ {G} - n _ {A} - n _ {G} - n _ {K} - n _ {L}\,,
\ee
we obtain
\be
E = 1 - L + I _ {G} - n _ {G} - n _ {K} - n _ {L} - 2 n _ {A}\,.
\ee
Employing the momentum conservation law for ghost and anti-ghost lines,
\be
2 I _ {G} - 2 n _ {G} = 2 n _ {L} + n _ {K} - E _ {c} -E _ {\bar{c}}\,,
\ee
we obtain
\be
E = 1 - L -\frac{1}{2} \left(n _ {K} + E _ {c} + E _ {\bar{c}}\right) - 2 n _ {A}\,.
\ee
It is clear that  the degree of divergence reduces as $n_{K}$, $E_c$ and $E_{\bar{c}}$ increase.
Thus, one may conclude the most divergent diagrams are 
those for which $n_{K}=E_c=E_{\bar{c}}=0$, i.e., the diagrams whose external lines are all gauge bosons. 
In this case, the degree of divergence is given by $E = 1 - 2 n _ {A} - L$. 
Since $n_ {A}$ is an integer, $E<0$, namely, 
the corresponding loop amplitudes ($ L \geq 1 $) are superficially convergent.
Here, we have utilized \textit{power-counting-in-exponentials} to understand the degree of divergence. 
In the local QFT limit of $M \to \infty$, all the exponential suppression factors disappear, 
and we will obtain divergences according to \textit{power-counting-in-polynomials} 
as usual in the standard QFT. 
We justify our procedure in the way that by taking $M \to \infty$, 
our results should reproduce the local theory results.

\subsection{Acknowledgments}
Authors thank Marco Frasca \& Alberto Salvio for useful comments.
This work is supported in part by the United States Department of Energy grant DE-SC0012447 
(N.O. and D.V.). AM’s research is financially supported by Netherlands Organisation for Scientific Research (NWO) grant number 680-91-11.

\bibliographystyle{apsrev4-1}

\begin{thebibliography}{99}

\medskip



\bibitem{sft1}
E.~Witten, ``{Noncommutative Geometry and String Field Theory},'' {\em
  Nucl.Phys.}, vol.~B268, p.~253, 1986.

\bibitem{sft2}
V.~A. Kostelecky and S.~Samuel, ``{On a Nonperturbative Vacuum for the Open
  Bosonic String},'' {\em Nucl.Phys.}, vol.~B336, p.~263, 1990.

\bibitem{sft3}
V.~A. Kostelecky and S.~Samuel, ``{The Static Tachyon Potential in the Open
  Bosonic String Theory},'' {\em Phys.Lett.}, vol.~B207, p.~169, 1988.

\bibitem{padic1}
P.~G. Freund and M.~Olson, ``{NONARCHIMEDEAN STRINGS},'' {\em Phys.Lett.},
  vol.~B199, p.~186, 1987.

\bibitem{padic2}
P.~G. Freund and E.~Witten, ``{ADELIC STRING AMPLITUDES},'' {\em Phys.Lett.},
  vol.~B199, p.~191, 1987.

\bibitem{padic3}
L.~Brekke, P.~G. Freund, M.~Olson, and E.~Witten, ``{Nonarchimedean String
  Dynamics},'' {\em Nucl.Phys.}, vol.~B302, p.~365, 1988.

\bibitem{Frampton-padic}
P.~H. Frampton and Y.~Okada, ``{Effective Scalar Field Theory of $P^-$adic
  String},'' {\em Phys.Rev.}, vol.~D37, pp.~3077--3079, 1988.

\bibitem{Tseytlin:1995uq} 
  A.~A.~Tseytlin,
  Phys.\ Lett.\ B {\bf 363}, 223 (1995)

\bibitem{marc}
T.~Biswas, M.~Grisaru, and W.~Siegel, ``{Linear Regge trajectories from
  worldsheet lattice parton field theory},'' {\em Nucl.Phys.}, vol.~B708,
  pp.~317--344, 2005.
  
  \bibitem{Siegel:2003vt} 
  W.~Siegel,
  ``Stringy gravity at short distances,''
  hep-th/0309093.
  

\bibitem{Biswas:2014yia} 
  T.~Biswas and N.~Okada,
  Nucl.\ Phys.\ B {\bf 898}, 113 (2015)
  doi:10.1016/j.nuclphysb.2015.06.023
  [arXiv:1407.3331 [hep-ph]].
  
  
\bibitem{Ghoshal:2017egr} 
  A.~Ghoshal, A.~Mazumdar, N.~Okada and D.~Villalba,
  Phys.\ Rev.\ D {\bf 97}, no. 7, 076011 (2018)
  doi:10.1103/PhysRevD.97.076011
  [arXiv:1709.09222 [hep-th]].


  
  

  
  
  \bibitem{Olive:2016xmw} 
  C.~Patrignani {\it et al.} [Particle Data Group],
  Chin.\ Phys.\ C {\bf 40}, no. 10, 100001 (2016).
  
  
  
\bibitem{Buoninfante:2018mre} 
  L.~Buoninfante, G.~Lambiase and A.~Mazumdar,
  arXiv:1805.03559 [hep-th].
  
  
\bibitem{Buoninfante:2018gce} 
  L.~Buoninfante, A.~Ghoshal, G.~Lambiase and A.~Mazumdar,
  arXiv:1812.01441 [hep-th].
  
  
\bibitem{Ghoshal:2018gpq}
A.~Ghoshal,
Int. J. Mod. Phys. A \textbf{34}, no.24, 1950130 (2019)
doi:10.1142/S0217751X19501306
[arXiv:1812.02314 [hep-ph]].


\bibitem{Frasca:2020jbe}
M.~Frasca and A.~Ghoshal,
[arXiv:2011.10586 [hep-th]].

\bibitem{Frasca:2021guf}
M.~Frasca and A.~Ghoshal,
[arXiv:2102.10665 [hep-th]].

  \bibitem{Biswas:2011ar} 
	 T.~Biswas, E.~Gerwick, T.~Koivisto and A.~Mazumdar,
  ``Towards singularity and ghost free theories of gravity,''
  Phys.\ Rev.\ Lett.\  {\bf 108}, 031101 (2012)
  doi:10.1103/PhysRevLett.108.031101
  [arXiv:1110.5249 [gr-qc]].
  
  
  \bibitem{Biswas:2013cha} 
  T.~Biswas, A.~Conroy, A.~S.~Koshelev and A.~Mazumdar,
  ``Generalized ghost-free quadratic curvature gravity,''
  Class.\ Quant.\ Grav.\  {\bf 31}, 015022 (2014)
  Erratum: [Class.\ Quant.\ Grav.\  {\bf 31}, 159501 (2014)]
  doi:10.1088/0264-9381/31/1/015022, 10.1088/0264-9381/31/15/159501
  [arXiv:1308.2319 [hep-th]].
  
  \bibitem{Frolov:2015bia} 
  V.~P.~Frolov, A.~Zelnikov and T.~de Paula Netto,
  ``Spherical collapse of small masses in the ghost-free gravity,''
  JHEP {\bf 1506}, 107 (2015)
  doi:10.1007/JHEP06(2015)107
  [arXiv:1504.00412 [hep-th]].
  
  
  \bibitem{Frolov:2015usa} 
  V.~P.~Frolov and A.~Zelnikov,
  ``Head-on collision of ultrarelativistic particles in ghost-free theories of gravity,''
  Phys.\ Rev.\ D {\bf 93}, no. 6, 064048 (2016)
  doi:10.1103/PhysRevD.93.064048
  [arXiv:1509.03336 [hep-th]].
  
 \bibitem{Koshelev:2018hpt} 
  A.~Koshelev, J.~Marto and A.~Mazumdar,
  ``Towards non-singular metric solution in infinite derivative gravity,''
  arXiv:1803.00309 [gr-qc].
  
\bibitem{Gama:2017ets}
F.~S.~Gama, J.~R.~Nascimento, A.~Y.~Petrov and P.~J.~Porfirio,
Phys. Rev. D \textbf{96}, no.10, 105009 (2017)
doi:10.1103/PhysRevD.96.105009
[arXiv:1710.02043 [hep-th]].

\bibitem{Gama:2020pte}
F.~S.~Gama, J.~R.~Nascimento and A.~Y.~Petrov,
Phys. Rev. D \textbf{101}, no.10, 105018 (2020)
doi:10.1103/PhysRevD.101.105018
[arXiv:2004.09299 [hep-th]].
  
  \bibitem{Koshelev:2017bxd} 
  A.~S.~Koshelev and A.~Mazumdar,
  ``Do massive compact objects without event horizon exist in infinite derivative gravity?,''
  Phys.\ Rev.\ D {\bf 96}, no. 8, 084069 (2017)
  doi:10.1103/PhysRevD.96.084069
  [arXiv:1707.00273 [gr-qc]].
  
 \bibitem{Buoninfante:2018xiw} 
  L.~Buoninfante, A.~S.~Koshelev, G.~Lambiase and A.~Mazumdar,
  ``Classical properties of non-local, ghost- and singularity-free gravity,''
  arXiv:1802.00399 [gr-qc].
  
  \bibitem{Cornell:2017irh} 
  A.~S.~Cornell, G.~Harmsen, G.~Lambiase and A.~Mazumdar,
  ``Rotating metric in Non-Singular Infinite Derivative Theories of Gravity,''
  arXiv:1710.02162 [gr-qc].
  
  \bibitem{Buoninfante:2018rlq} 
  L.~Buoninfante, A.~S.~Koshelev, G.~Lambiase, J.~Marto and A.~Mazumdar,
  "Conformally-flat, non-singular static metric in infinite derivative gravity,"
  arXiv:1804.08195 [gr-qc].
  
  \bibitem{Buoninfante:2018stt} 
  L.~Buoninfante, G.~Harmsen, S.~Maheshwari and A.~Mazumdar,
  "Non-singular metric for an electrically charged point-source in ghost-free infinite derivative gravity,"
  arXiv:1804.09624 [gr-qc].
  
   
\bibitem{Biswas:2005qr} 
  T.~Biswas, A.~Mazumdar and W.~Siegel,
  ``Bouncing universes in string-inspired gravity,''
  JCAP {\bf 0603}, 009 (2006)
  doi:10.1088/1475-7516/2006/03/009
  [hep-th/0508194].
  
  
  \bibitem{Biswas:2006bs} 
  T.~Biswas, R.~Brandenberger, A.~Mazumdar and W.~Siegel,
  ``Non-perturbative Gravity, Hagedorn Bounce and CMB,''
  JCAP {\bf 0712}, 011 (2007)
  doi:10.1088/1475-7516/2007/12/011
  [hep-th/0610274].
  
  
  \bibitem{Biswas:2010zk} 
  T.~Biswas, T.~Koivisto and A.~Mazumdar,
  ``Towards a resolution of the cosmological singularity in non-local higher derivative theories of gravity,''
  JCAP {\bf 1011}, 008 (2010)
  doi:10.1088/1475-7516/2010/11/008
  [arXiv:1005.0590 [hep-th]].

\bibitem{Biswas:2012bp} 
  T.~Biswas, A.~S.~Koshelev, A.~Mazumdar and S.~Y.~Vernov,
  ``Stable bounce and inflation in non-local higher derivative cosmology,''
  JCAP {\bf 1208}, 024 (2012)
  doi:10.1088/1475-7516/2012/08/024
  [arXiv:1206.6374 [astro-ph.CO]].
  
  
 \bibitem{Koshelev:2012qn} 
  A.~S.~Koshelev and S.~Y.~Vernov,
  ``On bouncing solutions in non-local gravity,''
  Phys.\ Part.\ Nucl.\  {\bf 43}, 666 (2012)
  doi:10.1134/S106377961205019X
  [arXiv:1202.1289 [hep-th]].
  
  \bibitem{Koshelev:2018rau} 
  A.~S.~Koshelev, J.~Marto and A.~Mazumdar,
  "Towards resolution of anisotropic cosmological singularity in infinite derivative gravity,"
  arXiv:1803.07072 [gr-qc].
  
  
  
  
\bibitem{Moffat:2011an} 
  J.~W.~Moffat,
  arXiv:1104.5706 [hep-th].
  
  \bibitem{Peskin}
  Introduction to QFT, Peskin and Schroder, Eqn. (16.46).
  
\bibitem{Talaganis:2014ida} 
  S.~Talaganis, T.~Biswas and A.~Mazumdar,
  Class.\ Quant.\ Grav.\  {\bf 32}, no. 21, 215017 (2015)
  doi:10.1088/0264-9381/32/21/215017
  [arXiv:1412.3467 [hep-th]].
  






  
  
  
  
\bibitem{Jegerlehner:2013nna} 
  F.~Jegerlehner,
  arXiv:1305.6652 [hep-ph].
  
  
  
  
\bibitem{EliasMiro:2012ay} 
  J.~Elias-Miro, J.~R.~Espinosa, G.~F.~Giudice, H.~M.~Lee and A.~Strumia,
  JHEP {\bf 1206}, 031 (2012)
  doi:10.1007/JHEP06(2012)031
  [arXiv:1203.0237 [hep-ph]].

\bibitem{Kadastik:2011aa} 
  M.~Kadastik, K.~Kannike, A.~Racioppi and M.~Raidal,
  JHEP {\bf 1205}, 061 (2012)
  doi:10.1007/JHEP05(2012)061
  [arXiv:1112.3647 [hep-ph]].

\bibitem{Gonderinger:2009jp} 
  M.~Gonderinger, Y.~Li, H.~Patel and M.~J.~Ramsey-Musolf,
  JHEP {\bf 1001}, 053 (2010)
  doi:10.1007/JHEP01(2010)053
  [arXiv:0910.3167 [hep-ph]].

  A.~Salvio,
  Phys.\ Lett.\ B {\bf 743}, 428 (2015)
  doi:10.1016/j.physletb.2015.03.015
  [arXiv:1501.03781 [hep-ph]].

\bibitem{Das:2015nwk} 
  A.~Das, N.~Okada and N.~Papapietro,
  Eur.\ Phys.\ J.\ C {\bf 77}, no. 2, 122 (2017)
  doi:10.1140/epjc/s10052-017-4683-2
  [arXiv:1509.01466 [hep-ph]].

\bibitem{Xiao:2014kba} 
  M.~L.~Xiao and J.~H.~Yu,
  Phys.\ Rev.\ D {\bf 90}, no. 1, 014007 (2014)
  Addendum: [Phys.\ Rev.\ D {\bf 90}, no. 1, 019901 (2014)]
  doi:10.1103/PhysRevD.90.014007, 10.1103/PhysRevD.90.019901
  [arXiv:1404.0681 [hep-ph]].
  


\bibitem{random1}
M.~R. Douglas and S.~H. Shenker, ``{Strings in Less Than One-Dimension},'' {\em
  Nucl.Phys.}, vol.~B335, p.~635, 1990.

\bibitem{random2}
D.~J. Gross and A.~A. Migdal, ``{Nonperturbative Solution of the Ising Model on
  a Random Surface},'' {\em Phys.Rev.Lett.}, vol.~64, p.~717, 1990.

\bibitem{random3}
E.~Brezin and V.~Kazakov, ``{Exactly Solvable Field Theories of Closed
  Strings},'' {\em Phys.Lett.}, vol.~B236, pp.~144--150, 1990.

\bibitem{ghoshal}
D.~Ghoshal, ``{p-adic string theories provide lattice discretization to the
  ordinary string worldsheet},'' {\em Phys.Rev.Lett.}, vol.~97, p.~151601,
  2006.

\bibitem{ncft1}
M.~R. Douglas and N.~A. Nekrasov, ``{Noncommutative field theory},'' {\em
  Rev.Mod.Phys.}, vol.~73, pp.~977--1029, 2001.

\bibitem{ncft2}
R.~J. Szabo, ``{Quantum field theory on noncommutative spaces},'' {\em
  Phys.Rept.}, vol.~378, pp.~207--299, 2003.

\bibitem{minimal}
S.~Hossenfelder, ``{Self-consistency in theories with a minimal length},'' {\em
  Class.Quant.Grav.}, vol.~23, pp.~1815--1821, 2006.

\bibitem{kdv}
I.~R.~A. Ludu, A. and W.~Greiner, ``{Generalized KdV Equation for Fluid
  Dynamics and Quantum Algebras},'' {\em Found. Phys.}, vol.~26, p.~665, 1996.

\bibitem{tachyon}
G.~Calcagni and G.~Nardelli, ``{Tachyon solutions in boundary and cubic string
  field theory},'' {\em Phys.Rev.}, vol.~D78, p.~126010, 2008.

\bibitem{sen-tachyon}
D.~Ghoshal and A.~Sen, ``{Tachyon condensation and brane descent relations in
  p-adic string theory},'' {\em Nucl.Phys.}, vol.~B584, pp.~300--312, 2000.

\bibitem{zwiebach}
N.~Moeller and B.~Zwiebach, ``{Dynamics with infinitely many time derivatives
  and rolling tachyons},'' {\em JHEP}, vol.~0210, p.~034, 2002.

\bibitem{minahan}
J.~A. Minahan, ``{Mode interactions of the tachyon condensate in p-adic string
  theory},'' {\em JHEP}, vol.~0103, p.~028, 2001.

\bibitem{BCK-solitons}
T.~Biswas, J.~A. Cembranos, and J.~I. Kapusta, ``{Finite Temperature Solitons
  in Non-Local Field Theories from p-Adic Strings},'' {\em Phys.Rev.},
  vol.~D82, p.~085028, 2010.

\bibitem{BCK}
T.~Biswas, J.~A. Cembranos, and J.~I. Kapusta, ``{Thermodynamics and
  Cosmological Constant of Non-Local Field Theories from p-Adic Strings},''
  {\em JHEP}, vol.~1010, p.~048, 2010.

\bibitem{Bluhm}
R.~Bluhm, ``{Particle fields at finite temperature from string field theory},''
  {\em Phys.Rev.}, vol.~D43, pp.~4042--4050, 1991.

\bibitem{BKR}
T.~Biswas, J.~Kapusta, and A.~Reddy, ``{Thermodynamics of String Field Theory
  Motivated Nonlocal Models},'' 2012.

\bibitem{BCK-solitons}
T.~Biswas, J.~A. Cembranos, and J.~I. Kapusta, ``{Finite Temperature Solitons
  in Non-Local Field Theories from p-Adic Strings},'' {\em Phys.Rev.},
  vol.~D82, p.~085028, 2010.

\bibitem{BCK}
T.~Biswas, J.~A. Cembranos, and J.~I. Kapusta, ``{Thermodynamics and
  Cosmological Constant of Non-Local Field Theories from p-Adic Strings},''
  {\em JHEP}, vol.~1010, p.~048, 2010.

\bibitem{Bluhm}
R.~Bluhm, ``{Particle fields at finite temperature from string field theory},''
  {\em Phys.Rev.}, vol.~D43, pp.~4042--4050, 1991.

\bibitem{BKR}
T.~Biswas, J.~Kapusta, and A.~Reddy, ``{Thermodynamics of String Field Theory
  Motivated Nonlocal Models},'' 2012.


\bibitem{Graham:2015cka} 
  P.~W.~Graham, D.~E.~Kaplan and S.~Rajendran,
  Phys.\ Rev.\ Lett.\  {\bf 115}, no. 22, 221801 (2015)
  doi:10.1103/PhysRevLett.115.221801
  [arXiv:1504.07551 [hep-ph]].

\bibitem{Pelaggi:2017wzr} 
  G.~M.~Pelaggi, F.~Sannino, A.~Strumia and E.~Vigiani,
  arXiv:1701.01453 [hep-ph].

\bibitem{Giudice:2016yja} 
  G.~F.~Giudice and M.~McCullough,
  JHEP {\bf 1702}, 036 (2017)
  doi:10.1007/JHEP02(2017)036
  [arXiv:1610.07962 [hep-ph]].

  
\bibitem{AADN}
A.~Ghoshal, A.~Mazumdar, D.~Villalba and N.~Okada
[draft in preparation]

\bibitem{Martin:2000cr}
S.~P.~Martin and J.~D.~Wells,
Phys. Rev. D \textbf{64}, 036010 (2001)
doi:10.1103/PhysRevD.64.036010
[arXiv:hep-ph/0011382 [hep-ph]].

\bibitem{Bond:2017suy}
A.~D.~Bond and D.~F.~Litim,
Phys. Rev. Lett. \textbf{119}, no.21, 211601 (2017)
doi:10.1103/PhysRevLett.119.211601
[arXiv:1709.06953 [hep-th]].



\bibitem{Adler:1982ri}
  S.~L.~Adler,
  ``Einstein Gravity as a Symmetry Breaking Effect in Quantum Field Theory,''
  Rev.\ Mod.\ Phys.\  {\bf 54} (1982) 729
   Erratum: [Rev.\ Mod.\ Phys.\  {\bf 55} (1983) 837].
  
\bibitem{Coleman:1973jx}
  S.~R.~Coleman and E.~J.~Weinberg,
  ``Radiative Corrections as the Origin of Spontaneous Symmetry Breaking,''
  Phys.\ Rev.\ D {\bf 7} (1973) 1888.
  
 

\bibitem{Salvio:2014soa}
  A.~Salvio and A.~Strumia,
  ``Agravity,''
  JHEP {\bf 1406} (2014) 080

\bibitem{Einhorn:2014gfa}
  M.~B.~Einhorn and D.~R.~T.~Jones,
  ``Naturalness and Dimensional Transmutation in Classically Scale-Invariant Gravity,''
  JHEP {\bf 1503} (2015) 047
  
  
\bibitem{Einhorn:2015lzy}
  M.~B.~Einhorn and D.~R.~T.~Jones,
  ``Induced Gravity I: Real Scalar Field,''
  JHEP {\bf 1601} (2016) 019
  
  
\bibitem{Einhorn:2016mws}
  M.~B.~Einhorn and D.~R.~T.~Jones,
  ``Induced Gravity II: Grand Unification,''
  JHEP {\bf 1605} (2016) 185
  
  
\bibitem{Khoze:2013uia}
  V.~V.~Khoze,
  ``Inflation and Dark Matter in the Higgs Portal of Classically Scale Invariant Standard Model,''
  JHEP {\bf 1311} (2013) 215
 
  
\bibitem{Kannike:2014mia}
  K.~Kannike, A.~Racioppi and M.~Raidal,
  ``Embedding inflation into the Standard Model - more evidence for classical scale invariance,''
  JHEP {\bf 1406} (2014) 154
  
\bibitem{Rinaldi:2014gha}
  M.~Rinaldi, G.~Cognola, L.~Vanzo and S.~Zerbini,
  ``Inflation in scale-invariant theories of gravity,''
  Phys.\ Rev.\ D {\bf 91} (2015) no.12,  123527
  
\bibitem{Kannike:2015apa}
  K.~Kannike, G.~Hutsi, L.~Pizza, A.~Racioppi, M.~Raidal, A.~Salvio and A.~Strumia,
  ``Dynamically Induced Planck Scale and Inflation,''
  JHEP {\bf 1505} (2015) 065
  
\bibitem{Kannike:2015fom}
  K.~Kannike, G.~Hutsi, L.~Pizza, A.~Racioppi, M.~Raidal, A.~Salvio and A.~Strumia,
  ``Dynamically Induced Planck Scale and Inflation,''
  PoS EPS {\bf -HEP2015} (2015) 379.
  
\bibitem{Barrie:2016rnv}
  N.~D.~Barrie, A.~Kobakhidze and S.~Liang,
  ``Natural Inflation with Hidden Scale Invariance,''
  Phys.\ Lett.\ B {\bf 756} (2016) 390
 
\bibitem{Tambalo:2016eqr}
  G.~Tambalo and M.~Rinaldi,
  ``Inflation and reheating in scale-invariant scalar-tensor gravity,''
  Gen.\ Rel.\ Grav.\  {\bf 49} (2017) no.4,  52
 
\bibitem{Hambye:2013sna}
  T.~Hambye and A.~Strumia,
  ``Dynamical generation of the weak and Dark Matter scale,''
  Phys.\ Rev.\ D {\bf 88} (2013) 055022

\bibitem{Karam:2015jta}
  A.~Karam and K.~Tamvakis,
  ``Dark matter and neutrino masses from a scale-invariant multi-Higgs portal,''
  Phys.\ Rev.\ D {\bf 92} (2015) no.7,  075010
  
\bibitem{Kannike:2016bny}
  K.~Kannike, G.~M.~Pelaggi, A.~Salvio and A.~Strumia,
  ``The Higgs of the Higgs and the diphoton channel,''
  JHEP {\bf 1607} (2016) 101
  
\bibitem{Hiller:2019mou}
G.~Hiller, C.~Hormigos-Feliu, D.~F.~Litim and T.~Steudtner,
Phys. Rev. D \textbf{102}, no.7, 071901 (2020)
doi:10.1103/PhysRevD.102.071901
[arXiv:1910.14062 [hep-ph]].

\bibitem{Hiller:2020fbu}
G.~Hiller, C.~Hormigos-Feliu, D.~F.~Litim and T.~Steudtner,
[arXiv:2008.08606 [hep-ph]].
  
\bibitem{Karam:2016rsz}
  A.~Karam and K.~Tamvakis,
  ``Dark Matter from a Classically Scale-Invariant $SU(3)_X$,''
  Phys.\ Rev.\ D {\bf 94} (2016) no.5,  055004
  

\bibitem{Bardeen}
  W. Bardeen, 

\bibitem{Foot:2007iy}
  R.~Foot, A.~Kobakhidze, K.~L.~McDonald and R.~R.~Volkas,
  ``A Solution to the hierarchy problem from an almost decoupled hidden sector within a classically scale invariant theory,''
  Phys.\ Rev.\ D {\bf 77} (2008) 035006
  
\bibitem{AlexanderNunneley:2010nw}
  L.~Alexander-Nunneley and A.~Pilaftsis,
  ``The Minimal Scale Invariant Extension of the Standard Model,''
  JHEP {\bf 1009} (2010) 021
  
\bibitem{Englert:2013gz}
  C.~Englert, J.~Jaeckel, V.~V.~Khoze and M.~Spannowsky,
  ``Emergence of the Electroweak Scale through the Higgs Portal,''
  JHEP {\bf 1304} (2013) 060
    
\bibitem{Farzinnia:2013pga}
  A.~Farzinnia, H.~J.~He and J.~Ren,
  ``Natural Electroweak Symmetry Breaking from Scale Invariant Higgs Mechanism,''
  Phys.\ Lett.\ B {\bf 727} (2013) 141
  
\bibitem{Holthausen:2013ota}
  M.~Holthausen, J.~Kubo, K.~S.~Lim and M.~Lindner,
  ``Electroweak and Conformal Symmetry Breaking by a Strongly Coupled Hidden Sector,''
  JHEP {\bf 1312} (2013) 076
  
\bibitem{Altmannshofer:2014vra}
  W.~Altmannshofer, W.~A.~Bardeen, M.~Bauer, M.~Carena and J.~D.~Lykken,
  ``Light Dark Matter, Naturalness, and the Radiative Origin of the Electroweak Scale,''
  JHEP {\bf 1501} (2015) 032
  
\bibitem{Farzinnia:2015fka}
  A.~Farzinnia and S.~Kouwn,
 ``Classically scale invariant inflation, supermassive WIMPs, and adimensional gravity,''
  Phys.\ Rev.\ D {\bf 93} (2016) no.6,  063528
  doi:10.1103/PhysRevD.93.063528
  
\bibitem{Tavares:2013dga}
G.~Marques Tavares, M.~Schmaltz and W.~Skiba,
Phys. Rev. D \textbf{89}, no.1, 015009 (2014)
doi:10.1103/PhysRevD.89.015009
[arXiv:1308.0025 [hep-ph]].

\bibitem{Pelaggi:2017abg}
G.~M.~Pelaggi, A.~D.~Plascencia, A.~Salvio, F.~Sannino, J.~Smirnov and A.~Strumia,
Phys. Rev. D \textbf{97}, no.9, 095013 (2018)
doi:10.1103/PhysRevD.97.095013
[arXiv:1708.00437 [hep-ph]].

\bibitem{Giudice:2014tma}
G.~F.~Giudice, G.~Isidori, A.~Salvio and A.~Strumia,
JHEP \textbf{02}, 137 (2015)
doi:10.1007/JHEP02(2015)137
[arXiv:1412.2769 [hep-ph]].

\cite{Litim:2014uca}
\bibitem{Litim:2014uca}
D.~F.~Litim and F.~Sannino,
JHEP \textbf{12}, 178 (2014)
doi:10.1007/JHEP12(2014)178
[arXiv:1406.2337 [hep-th]].


\bibitem{Litim:2015iea}
D.~F.~Litim, M.~Mojaza and F.~Sannino,
JHEP \textbf{01}, 081 (2016)
doi:10.1007/JHEP01(2016)081
[arXiv:1501.03061 [hep-th]].



\bibitem{Molgaard:2016bqf}
Accepted for publication in Physical Review D. 


\bibitem{Abel:2017ujy}
S.~Abel and F.~Sannino,
Phys. Rev. D \textbf{96}, no.5, 056028 (2017)
doi:10.1103/PhysRevD.96.056028
[arXiv:1704.00700 [hep-ph]].


\bibitem{Esbensen:2015cjw}
J.~K.~Esbensen, T.~A.~Ryttov and F.~Sannino,
Phys. Rev. D \textbf{93}, no.4, 045009 (2016)
doi:10.1103/PhysRevD.93.045009
[arXiv:1512.04402 [hep-th]].


\bibitem{Bond:2017lnq}
A.~D.~Bond and D.~F.~Litim,
Phys. Rev. D \textbf{97}, no.8, 085008 (2018)
doi:10.1103/PhysRevD.97.085008
[arXiv:1707.04217 [hep-th]].


\bibitem{Abel:2017rwl}
S.~Abel and F.~Sannino,
Phys. Rev. D \textbf{96}, no.5, 055021 (2017)
doi:10.1103/PhysRevD.96.055021
[arXiv:1707.06638 [hep-ph]].


\bibitem{Intriligator:2015xxa}
K.~Intriligator and F.~Sannino,
JHEP \textbf{11}, 023 (2015)
doi:10.1007/JHEP11(2015)023
[arXiv:1508.07411 [hep-th]].
%
%
%


\bibitem{Bajc:2016efj}
B.~Bajc and F.~Sannino,
JHEP \textbf{12}, 141 (2016)
doi:10.1007/JHEP12(2016)141
[arXiv:1610.09681 [hep-th]].


\bibitem{SUSY}
See e.g.\ the talks at the ``
Madrid, 28-30 September 2016.


\bibitem{Holdom:2010qs}
B.~Holdom,
Phys. Lett. B \textbf{694}, 74-79 (2011)
doi:10.1016/j.physletb.2010.09.037
[arXiv:1006.2119 [hep-ph]].


\bibitem{Pica:2010xq}
C.~Pica and F.~Sannino,
Phys. Rev. D \textbf{83}, 035013 (2011)
doi:10.1103/PhysRevD.83.035013
[arXiv:1011.5917 [hep-ph]].


\bibitem{Shrock:2013cca}
R.~Shrock,
Phys. Rev. D \textbf{89}, no.4, 045019 (2014)
doi:10.1103/PhysRevD.89.045019
[arXiv:1311.5268 [hep-th]].


\bibitem{PalanquesMestre:1983zy}
A.~Palanques-Mestre and P.~Pascual,
Commun. Math. Phys. \textbf{95}, 277 (1984)
doi:10.1007/BF01212398




\bibitem{Gracey:1996he}
J.~Gracey,
Phys. Lett. B \textbf{373}, 178-184 (1996)
[arXiv:hep-ph/9602214 [hep-ph]].


\bibitem{Bond:2017wut}
A.~D.~Bond, G.~Hiller, K.~Kowalska and D.~F.~Litim,
JHEP \textbf{08}, 004 (2017)
[arXiv:1702.01727 [hep-ph]].
 
 
\bibitem{Kowalska:2017pkt}
K.~Kowalska and E.~M.~Sessolo,
JHEP \textbf{04}, 027 (2018)
[arXiv:1712.06859 [hep-ph]].


\bibitem{Mann:2017wzh}
R.~Mann, J.~Meffe, F.~Sannino, T.~Steele, Z.~W.~Wang and C.~Zhang,
Phys. Rev. Lett. \textbf{119}, no.26, 261802 (2017)
[arXiv:1707.02942 [hep-ph]].



\bibitem{Pelaggi:2017abg}
G.~M.~Pelaggi, A.~D.~Plascencia, A.~Salvio, F.~Sannino, J.~Smirnov and A.~Strumia,
Phys. Rev. D \textbf{97}, no.9, 095013 (2018)
[arXiv:1708.00437 [hep-ph]].

\bibitem{Burgess:2009ea}
C.~Burgess, H.~M.~Lee and M.~Trott,
JHEP \textbf{09}, 103 (2009)
[arXiv:0902.4465 [hep-ph]].

\bibitem{Barbon:2009ya}
J.~Barbon and J.~Espinosa,
Phys. Rev. D \textbf{79}, 081302 (2009)
[arXiv:0903.0355 [hep-ph]].

\bibitem{Bezrukov:2010jz}
F.~Bezrukov, A.~Magnin, M.~Shaposhnikov and S.~Sibiryakov,
JHEP \textbf{01}, 016 (2011)
[arXiv:1008.5157 [hep-ph]].


\bibitem{Cornwall:1974km}
J.~M.~Cornwall, D.~N.~Levin and G.~Tiktopoulos,
Phys. Rev. D \textbf{10}, 1145 (1974)


\bibitem{Shaposhnikov:2018xkv}
M.~Shaposhnikov and A.~Shkerin,
Phys. Lett. B \textbf{783}, 253-262 (2018)
[arXiv:1803.08907 [hep-th]].

\bibitem{Shaposhnikov:2018jag}
M.~Shaposhnikov and A.~Shkerin,
JHEP \textbf{10}, 024 (2018)
[arXiv:1804.06376 [hep-th]].

\bibitem{Mooij:2018hew}
S.~Mooij, M.~Shaposhnikov and T.~Voumard,
Phys. Rev. D \textbf{99}, no.8, 085013 (2019)
[arXiv:1812.07946 [hep-th]].

\bibitem{Shaposhnikov:2018nnm}
M.~Shaposhnikov and K.~Shimada,
Phys. Rev. D \textbf{99}, no.10, 103528 (2019)
[arXiv:1812.08706 [hep-ph]].









\end{thebibliography}

\end{document}